%% file: main_arxiv.tex
\documentclass[twocolumn]{fairmeta}

\newif\ifmlsystemplate
\mlsystemplatefalse

\hypersetup{
    pdftitle={Agentic Operator Generation for ML ASICs},
    pdfsubject={Machine Learning Systems, Compiler Optimization, Hardware Acceleration},
    pdfkeywords={Triton, MTIA, PyTorch, Kernel Generation, LLM, Compiler, ASIC},
    pdfproducer={LaTeX with fairmeta class},
    breaklinks=true,
    colorlinks=true,
    linkcolor=metablue,
    citecolor=metablue,
    urlcolor=metablue,
    pdfborder={0 0 0},
    citebordercolor={0 0 0},
    linkbordercolor={0 0 0},
    urlbordercolor={0 0 0}
}

\usepackage{pifont}  
\usepackage{listings}  
\usepackage{xspace}
\usepackage{inconsolata}
\usepackage{siunitx}
\usepackage{algorithm}
\usepackage{algpseudocode}

\newcolumntype{L}{>{\ttfamily\small}l}
\newcommand{\op}[1]{\texttt{\nolinkurl{#1}}}

\usepackage{xcolor}
\definecolor{codegray}{gray}{0.95}
\definecolor{codepurple}{rgb}{0.58,0,0.82}
\definecolor{codeblue}{rgb}{0.25,0.5,0.75}
\definecolor{codegreen}{rgb}{0,0.6,0}
\usepackage{url}

\lstdefinestyle{pythonstyle}{
    language=Python,
    backgroundcolor=\color{codegray},
    basicstyle=\ttfamily\small,
    keywordstyle=\color{codeblue}\bfseries,
    commentstyle=\color{codegreen}\itshape,
    stringstyle=\color{codepurple},
    numbers=left,
    numberstyle=\tiny,
    stepnumber=1,
    numbersep=8pt,
    showstringspaces=false,
    breaklines=true,
    tabsize=4,
    frame=single,
    captionpos=b
}

\definecolor{cwm_keyword_color}{HTML}{0064E0}
\definecolor{cwm_string_color}{HTML}{009B9B}
\definecolor{cwm_comment_color}{HTML}{D75FAA}

\lstdefinestyle{simple}{
  basicstyle=\ttfamily\small,
  columns=fullflexible,
  backgroundcolor=\color{gray!8},
  frame=single,
  rulecolor=\color{black!30},
  numberstyle=\tiny\color{gray},
  keywordstyle=\color{blue},
  commentstyle=\color{green!40!black},
  stringstyle=\color{red!70!black},
  showstringspaces=false,
  tabsize=2,
  breaklines=true,
  breakatwhitespace=true,
  captionpos=b,
  xleftmargin=3.4pt,
  xrightmargin=3.4pt
}

\lstdefinestyle{simple_small}{
  basicstyle=\fontsize{8}{10}\ttfamily,
  columns=fullflexible,
  keywordstyle=\color{cwm_keyword_color},
  commentstyle=\color{cwm_comment_color},
  stringstyle=\color{cwm_string_color},
  breaklines=true,
  frame=single,
  escapeinside={(*@}{@*)}
}

\lstdefinestyle{simple_small_small}{
  basicstyle=\fontsize{5}{6}\ttfamily,
  columns=fullflexible,
  keywordstyle=\color{cwm_keyword_color},
  commentstyle=\color{cwm_comment_color},
  stringstyle=\color{cwm_string_color},
  breaklines=true,
  frame=single,
  escapeinside={(*@}{@*)}
}

\definecolor{cwm_think_color}{HTML}{AFD7FF}
\lstdefinestyle{cwm_think}{
  framerule=0pt,
  columns=fullflexible,
  basicstyle=\fontsize{5}{6}\ttfamily,
  backgroundcolor=\color{cwm_think_color},
  showstringspaces=false,
  breaklines=true,
  xleftmargin=3.4pt,
  xrightmargin=3.4pt,
  breakindent=0pt
}

\definecolor{cwm_prompt_color}{HTML}{A8E6CF}
\lstdefinestyle{cwm_prompt}{
  framerule=0pt,
  columns=fullflexible,
  basicstyle=\fontsize{5}{6}\ttfamily,
  backgroundcolor=\color{cwm_prompt_color},
  showstringspaces=false,
  breaklines=true,
  xleftmargin=3.4pt,
  xrightmargin=3.4pt,
  breakindent=0pt
}

\definecolor{cwm_act_color}{HTML}{D2D2FF}
\lstdefinestyle{cwm_act}{
  framerule=0pt,
  columns=fullflexible,
  basicstyle=\fontsize{5}{6}\ttfamily,
  backgroundcolor=\color{cwm_act_color},
  showstringspaces=false,
  breaklines=true,
  xleftmargin=3.4pt,
  xrightmargin=3.4pt,
  breakindent=0pt
}

\definecolor{cwm_obs_color}{HTML}{FFDCB9}
\lstdefinestyle{cwm_obs}{
  framerule=0pt,
  columns=fullflexible,
  keepspaces=true,
  basicstyle=\fontsize{5}{6}\ttfamily,
  backgroundcolor=\color{cwm_obs_color},
  showstringspaces=false,
  breaklines=true,
  xleftmargin=3.4pt,
  xrightmargin=3.4pt,
  breakindent=0pt
}

\input{preamble_common}

\input{sections/00_decl}

\title{Agentic Operator Generation for ML ASICs}

\author[1, *]{Alec M. Hammond}
\author[2, *]{Aram Markosyan}
\author[1]{Aman Dontula}
\author[1]{Simon Mahns}
\author[2]{Zacharias Fisches}
\author[2]{Dmitrii Pedchenko}
\author[2]{Keyur Muzumdar}
\author[2]{Natacha Supper}
\author[3]{Mark Saroufim}
\author[3]{Joe Isaacson}
\author[3]{Laura Wang}
\author[2]{Warren Hunt}
\author[1]{Kaustubh Gondkar}
\author[1]{Roman Levenstein}
\author[2]{Gabriel Synnaeve}
\author[1]{Richard Li}
\author[2]{Jacob Kahn}
\author[1]{Ajit Mathews}

\correspondence{Alec M. Hammond (\email{alechammond@meta.com}), Jacob Kahn (\email{jacobkahn@meta.com})}

\affiliation[1]{Meta}
\affiliation[2]{FAIR, Meta Superintelligence Labs}
\affiliation[3]{Meta Superintelligence Labs}

\contribution[*]{Equal contribution.}

\date{\today}

\abstract{
We present \agentname, an agentic AI system designed to generate functionally correct Triton PyTorch \aten{} kernels at scale for emerging accelerator platforms. \agentname{} integrates open-source large language models with a custom linter, JIT compilation, and a PyTorch \opinfo{}-based test harness. This pipeline is compatible with both real Meta Training and Inference Accelerator (MTIA) silicon and in hardware simulation environments for next-generation devices.
In contrast to previous kernel-generation approaches that prioritize performance for a limited set of high-usage kernels, \agentname{} prioritizes coverage. Our system emphasizes correctness and generality across the entire operator set, including diverse data types, shapes, and argument patterns. In our experiments, \agentname{} successfully generated kernels and wrappers for 481 unique \aten{} operators that pass all corresponding PyTorch \opinfo{} tests (over \totaltests{} in total).
\agentname{} paves the way for overnight generation of complete PyTorch \aten{} backends for new accelerator platforms.
}

\begin{document}
\maketitle

\input{sections/01_intro}
\input{sections/02_background}
\input{sections/03_system_design}
\input{sections/04_experiments_results}
\input{sections/06_related_work}
\input{sections/07_discussion}
\input{sections/08_conclusion}

\bibliographystyle{assets/plainnat}
\bibliography{main}

\clearpage
\onecolumn
\beginappendix
\input{sections/99_appendix}

\end{document}

%% file: preamble_common.tex
\newcommand{\agentname}{TritorX}
\newcommand{\opinfo}{OpInfo}
\newcommand{\aten}{ATen}
\newcommand{\totaltests}{20,000}

\providecommand{\appenref}[1]{\Cref{#1}}

\def\figref#1{Figure~\ref{#1}}  


%% file: sections/00_decl.tex
\ifmlsystemplate\else

\usepackage[utf8]{inputenc}
\usepackage{subcaption}

\usepackage[T1]{fontenc}    
\PassOptionsToPackage{hyphens}{url}
\usepackage{url}            
\usepackage{hyperref}       
\usepackage{booktabs}       
\usepackage{amsfonts}       
\usepackage{nicefrac}       
\usepackage{microtype}      
\usepackage[dvipsnames]{xcolor}

\usepackage{nicematrix}
\usepackage{algorithm}
\usepackage{algpseudocode}
\usepackage{amsthm}
\usepackage[normalem]{ulem}
\usepackage{wrapfig}
\usepackage{graphicx}
\usepackage{cancel}
\usepackage{multirow}
\usepackage{multicol}
\usepackage{amssymb}
\usepackage{cleveref}
\usepackage{colortbl}
\usepackage{xspace}
\usepackage{inconsolata}

\usepackage{graphicx}

\usepackage{xcolor,ifthen,marginnote}
\usepackage{listings}

\newboolean{showcomments}

\usepackage{etoolbox}

\crefformat{section}{\S#2#1#3}
\Crefformat{section}{\S#2#1#3}
\crefmultiformat{section}{\S#2#1#3}{ and \S#2#1#3}{, \S#2#1#3}{ and \S#2#1#3}
\Crefmultiformat{section}{\S#2#1#3}{ and \S#2#1#3}{, \S#2#1#3}{ and \S#2#1#3}
\crefrangeformat{section}{\S#3#1#4 to \S#5#2#6}
\Crefrangeformat{section}{\S#3#1#4 to \S#5#2#6}

\newtoggle{release}

\togglefalse{release}

\usepackage{siunitx}

\iftoggle{release}{
\newcommand{\jacob}[1]{}
\newcommand{\alec}[1]{}
\newcommand{\zach}[1]{}
\newcommand{\kmuzumdar}[1]{}
\newcommand{\openrichard}[1]{}
}
{
\newcommand{\jacob}[1]{\textcolor{green}{JK: #1}}
\newcommand{\alec}[1]{\textcolor{red}{AH: #1}}
\newcommand{\zach}[1]{\textcolor{orange!70!black}{ZF: #1}}
\newcommand{\kmuzumdar}[1]{\textcolor{gray}{KM: #1}}
\newcommand{\openrichard}[1]{\textcolor{magenta}{OR: #1}}
}

\ifdefined\mlsystemplatetrue
  \ifmlsystemplate
    \usepackage{draftwatermark}
    \SetWatermarkText{DRAFT}          
    \SetWatermarkScale{1.2}             
    \SetWatermarkLightness{0.95}      
    \SetWatermarkAngle{45}
  \fi
\fi

\lstdefinestyle{simple}{
  basicstyle=\ttfamily\small,     
  columns=fullflexible,
  backgroundcolor=\color{gray!8}, 
  frame=single,                    
  rulecolor=\color{black!30},      
  numberstyle=\tiny\color{gray},   
  keywordstyle=\color{blue},       
  commentstyle=\color{green!40!black}, 
  stringstyle=\color{red!70!black},    
  showstringspaces=false,          
  tabsize=2,                       
  breaklines=true,                 
  breakatwhitespace=true,          
  captionpos=b,                     
  xleftmargin=3.4pt,
  xrightmargin=3.4pt
}

\definecolor{cwm_keyword_color}{HTML}{0064E0}
\definecolor{cwm_string_color}{HTML}{009B9B}
\definecolor{cwm_string_color_bright}{HTML}{C80A28}
\definecolor{cwm_comment_color}{HTML}{D75FAA}

\lstdefinestyle{simple_small}{
  basicstyle=\fontsize{8}{10}\ttfamily, 
  columns=fullflexible,
  keywordstyle=\color{cwm_keyword_color},
  commentstyle=\color{cwm_comment_color},
  stringstyle=\color{cwm_string_color},
  breaklines=true,
  frame=single,
  escapeinside={(*@}{@*)}
}

\lstdefinestyle{simple_small_small}{
  basicstyle=\fontsize{5}{6}\ttfamily, 
  columns=fullflexible,
  keywordstyle=\color{cwm_keyword_color},
  commentstyle=\color{cwm_comment_color},
  stringstyle=\color{cwm_string_color},
  breaklines=true,
  frame=single,
  escapeinside={(*@}{@*)}
}

\definecolor{cwm_think_color}{HTML}{AFD7FF}
\lstdefinestyle{cwm_think}{
  framerule=0pt,
  columns=fullflexible,
  basicstyle=\fontsize{5}{6}\ttfamily,     
  backgroundcolor=\color{cwm_think_color}, 
  showstringspaces=false,          
  breaklines=true,                 
  xleftmargin=3.4pt,
  xrightmargin=3.4pt,
  breakindent=0pt
}
\definecolor{cwm_prompt_color}{HTML}{A8E6CF}
\lstdefinestyle{cwm_prompt}{
  framerule=0pt,
  columns=fullflexible,
  basicstyle=\fontsize{5}{6}\ttfamily,     
  backgroundcolor=\color{cwm_prompt_color}, 
  showstringspaces=false,          
  breaklines=true,                 
  xleftmargin=3.4pt,
  xrightmargin=3.4pt,
  breakindent=0pt
}

\definecolor{cwm_act_color}{HTML}{D2D2FF}
\lstdefinestyle{cwm_act}{
  framerule=0pt,
  columns=fullflexible,
  basicstyle=\fontsize{5}{6}\ttfamily,     
  backgroundcolor=\color{cwm_act_color}, 
  showstringspaces=false,          
  breaklines=true,                 
  xleftmargin=3.4pt,
  xrightmargin=3.4pt,
  breakindent=0pt
}

\definecolor{cwm_obs_color}{HTML}{FFDCB9}
\lstdefinestyle{cwm_obs}{
  framerule=0pt,
  columns=fullflexible,
  keepspaces=true,
  basicstyle=\fontsize{5}{6}\ttfamily,     
  backgroundcolor=\color{cwm_obs_color}, 
  showstringspaces=false,          
  breaklines=true,                 
  xleftmargin=3.4pt,
  xrightmargin=3.4pt,
  breakindent=0pt
}

\usepackage{placeins}

%% file: sections/01_intro.tex
\vspace{-0.1cm}
\section{Introduction}
\label{sec:intro}

\begin{figure}[ht!]
    \centering
    \includegraphics[width=\columnwidth]{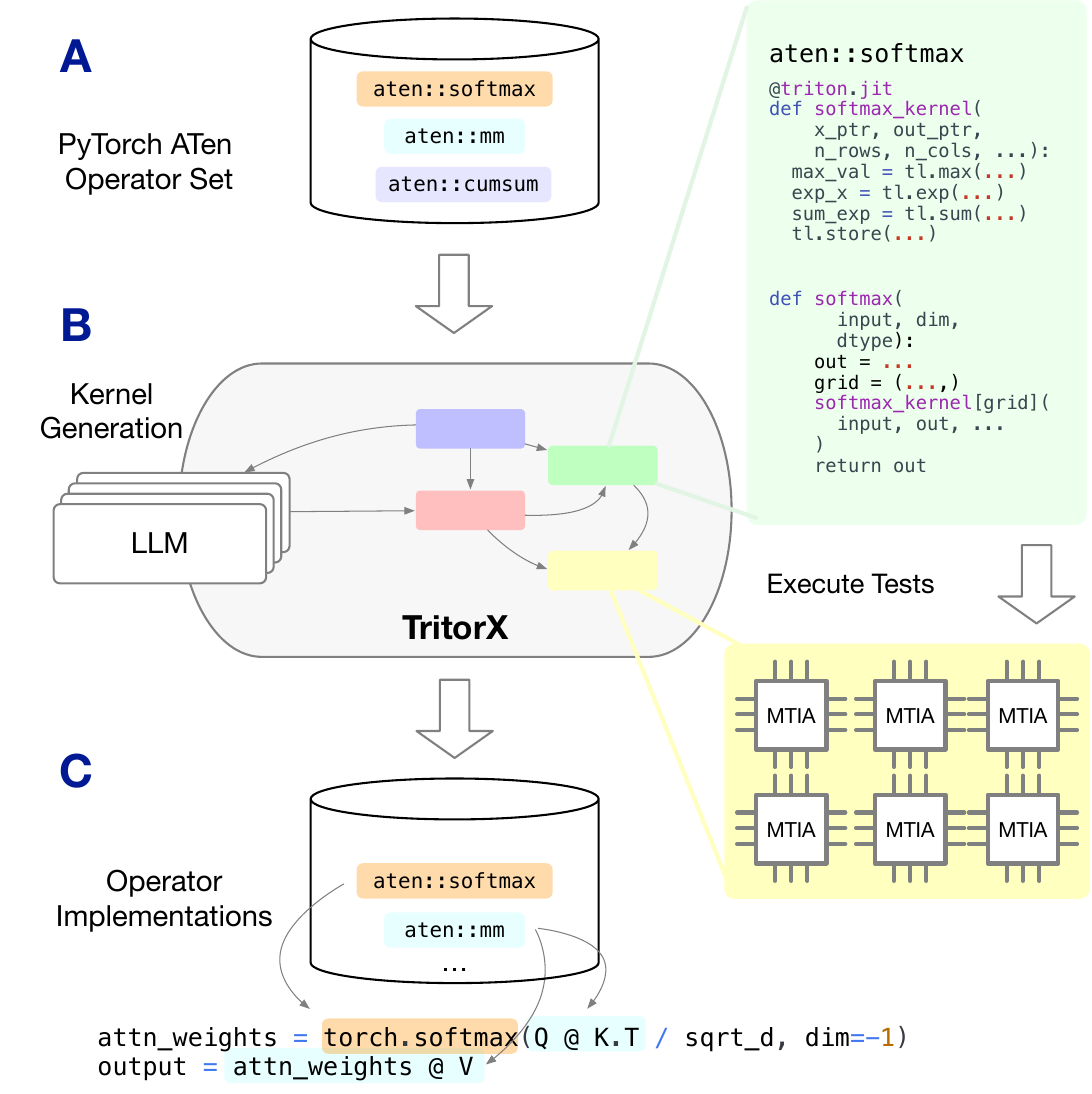}
    \caption{\emph{\agentname{} System Overview.} \textbf{(A)} \opinfo{} operators and their PyTorch docstring/signature are selected for generation. \textbf{(B)} \agentname{} iterates on each operator in a finite-state machine feedback loop. Kernel-wrapper pairs are generated using a large language model (LLM). Production infrastructure allows for simultaneous generation and testing at scale. \textbf{(C)} The generation task is successful if an operator passes all corresponding \opinfo{} tests.
    }
    \label{fig:system}
\end{figure}

The rapid adoption of machine learning (ML) and artificial intelligence (AI) hardware is projected to drive US datacenter power consumption to between $6.7\%$ and $12.0\%$ of total electricity usage by 2028~\citep{shehabi20242024}, highlighting an urgent need for efficient accelerator hardware to support large-scale model inference and training. In response, the industry is investing heavily in heterogeneous datacenter fleets that incorporate a variety of accelerator solutions, including custom silicon ASICs tailored to specific workloads and requirements~\citep{accelerators_survey}. Namely, the Meta Training and Inference Accelerator (MTIA) currently serves recommendation models (DLRM)~\citep{meta_dlrm} to billions of users across Facebook, Instagram, and Threads, while simultaneously reducing total cost of ownership by 44\% compared to GPUs~\citep{mtia_artemis}.

However, despite the obvious advantages offered by in-house accelerators, each new platform requires significant engineering labor to build a software ecosystem compatible with existing tools such as PyTorch~\citep{NEURIPS2019_bdbca288}, given the large set of required tensor operators~\citep{kahn2022flashlight}. An aspect therein is \emph{operator coverage}, or the fraction of operators in \aten~\citep{paszke2017automatic} --- PyTorch's tensor library --- that have kernels executing natively on a particular accelerator. Establishing and maintaining comprehensive operator coverage is an arduous task that is needed towards running inference with new or for prototyping new model architectures for training. In other words, while new accelerator platforms must provide competitive performance per unit cost, they must \emph{also} provide an amenable developer experience with a comprehensive kernel backend.

To address this challenge, we present an agentic AI system capable of generating functionally correct Triton ATen kernels for MTIA at scale. This new tool, which we call \agentname, leverages open-source large language models (LLM) paired with execution feedback in a finite state machine (FSM) to generate, compile, and test hundreds of kernels directly on MTIA hardware. \agentname{} is compatible with MTIA's production infrastructure, producing kernel-wrapper pairs that can be immediately registered within PyTorch and can be used for experimental model training, or in a production inference model. Building on top of production infrastructure, \agentname{} can also generate new PyTorch backends for upcoming accelerator generations via hardware simulation, providing important feedback to hardware and compiler engineers \emph{before} tape-out. In the limit, we envision implementing a kernel backend for a new chipset overnight.

We note that \agentname{} differs from other kernel generation efforts, in that our framework strictly optimizes for correctness and generalizability across an entire backend, rather than performance over a narrow subset of critical path kernels~\citep{aicuda}. For example, \agentname{} is configured to generate kernel-wrapper pairs that are compatible with a wide range of quantization data types, tensor shapes, or PyTorch argument inputs. In some cases, \agentname{} will generate multiple kernel implementations for a particular operator, and the corresponding dispatch logic is implemented in the wrapper function. Using a Triton dialect known as Triton MTIA, standard models can produce working kernels and rely on execution feedback to identify MTIA-specific semantics or intrinsics. Importantly, we only provide \agentname{} with the \aten{} operator docstring to generate the corresponding implementation. \figref{fig:system} describes the end-to-end workflow for accelerator enablement.

\agentname{} performs in-context learning iteratively, distilling hardware requirements and their corresponding Triton semantics based on the feedback obtained directly via tools like the compiler. Previous works demonstrated that pipelines like \agentname{} will often ``cheat'' at the generation task by dispatching to the host or calling other undefined PyTorch functions in the operator wrapper~\citep{sakana2025robustkbench}. We avoid this by incorporating a custom linter that catches unauthorized uses of these functions or utilities and forces correction via feedback.

Central to \agentname{} is the integration of multiple \emph{testing} frameworks to ensure correctness across different data types, tensor shapes, or input arguments. We use \opinfo\footnote{\hyperlink{https://github.com/pytorch/pytorch/blob/main/torch/testing/_internal/opinfo/core.py}{OpInfo in the PyTorch core.}}, a PyTorch-native testing framework, along with a custom test harness that pulls test data from models in production.

In summary, our contributions are as follows:
\begin{itemize}

  \item \textbf{System.} A finite state machine using open source LLMs that generates, compiles, and validates hundreds of Triton MTIA kernels end-to-end. \agentname{} runs directly on deployed silicon or via hardware simulation enabling prototyping for future devices. \agentname{} only requires operator docstrings to generate the corresponding implementations. Its modular design allows for modular features, like the custom linter which prevents ``cheating.''

  \item \textbf{Coverage objective.} A design that optimizes for functional correctness and backend coverage covering multiple data types, shapes, and branching dispatch logic via generated wrappers.

  \item \textbf{Results.} At scale, \agentname{} produced 481 ATen kernels (84.7\% MTIA-compatible \opinfo{} coverage) which pass all their corresponding \opinfo{} tests, in total more than \totaltests{}. In addition, \agentname{} demonstrates capability on end-to-end model enablement tasks for various models on MTIA. \agentname{} can iterate over the entire \opinfo{} operator set \emph{in a few hours}, allowing for rapid backend development and iteration.

  \item \textbf{Testing harness}. We integrate PyTorch \emph{OpInfo} and captured production inputs testing for correctness under deployment conditions.
\end{itemize}
\vspace{-0.1cm}

%% file: sections/02_background.tex
\section{Background}
\label{sec:background}

In this section, we describe the MTIA architecture, the Triton MTIA dialect, and how \agentname{} can be instrumented to generate kernels at scale.

\begin{figure}[ht]
    \centering
    \includegraphics[width=\linewidth]{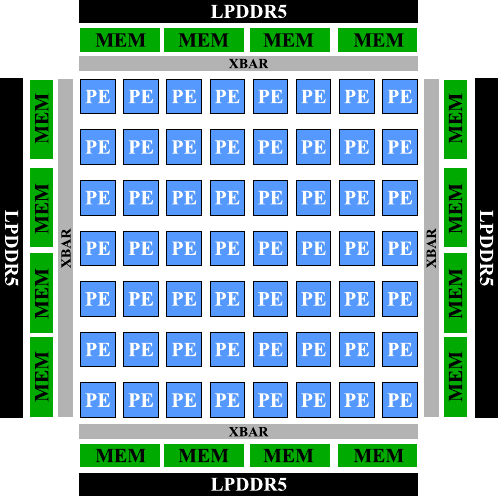}
    \caption{MTIA's architecture overview. The core kernel computation is performed by a grid of process elements (PEs) which consist of a scalar core, a vector core, and special function units.}
    \label{fig:mtia_arch}
\end{figure}

The MTIA architecture employs a grid of 8x8 processing elements (PEs) responsible for executing the core kernel workloads~\citep{mtia_artemis}. Each PE consists of a scalar RISC-V core, a vector RISC-V core, and various fixed-function units (FFUs) responsible for implementing dedicated computations, such as direct memory accesses (DMAs) and dot products. \figref{fig:mtia_arch} illustrates the MTIA architecture.

In particular, MTIA features a unique memory hierarchy with a significant amount of local SRAM available to all PEs via a series of crossbars. Dedicated PE FFUs facilitate the abstraction of circular buffers, enabling efficient pipelining of computation and communication needed to amortize the overall latency resulting from data movement. The computational savings gained by this approach allow MTIA to leverage cheaper LPDDR DRAM instead of HBM. Importantly, significant effort was made to mitigate overhead related to kernel or job dispatch from the host, enabling eager-mode workflows.

To execute kernels on MTIA, developers can author the workload using a C++ API, which is compiled with an LLVM backend, or using a custom Triton dialect adapted specifically for MTIA. Triton is an open-source Python library that provides a domain-specific language (DSL) for writing highly efficient custom GPU kernels~\citep{triton_paper}. It simplifies GPU programming by allowing developers to express complex parallel computations using intuitive block-based abstractions while automatically handling low-level details such as memory access patterns and synchronization. This enables users to achieve performance comparable to hand-written CUDA code, without requiring deep expertise in GPU architectures.

Although Triton was written to express the semantics of GPU computation, several of the existing semantics can be directly translated to corresponding MTIA hardware features. For example, instead of mapping Triton blocks to GPU threads, we can map them to the MTIA PE grid and rely on masking within loads and stores such that tensor boundaries are respected. Furthermore, loads and stores can take advantage of the MTIA DMA engine for structured memory access. When intrinsics do not perfectly match, we can \emph{augment} the underlying Triton feature-set with specific device libraries (e.g. to leverage FFUs that implement nonlinear activations). Triton MTIA is a dedicated dialect intended to facilitate the above mappings, along with various other performance optimizations via the compiler backend.

Although Triton MTIA intends to preserve the existing Triton semantics as much as possible, there are certain hardware requirements that force notable deviations. For example, MTIA requires 32-byte aligned memory access patterns, and load/store operations will fail if this is not satisfied. On the surface, this may seem problematic when trying to generate kernels using off-the-shelf models. However, Triton MTIA has detailed assert messages and error handling providing the necessary feedback that the models need to adapt the vanilla Triton code for MTIA. Indeed, building a compiler tool-chain that gives descriptive feedback is an important part of leveraging automated approaches for code generation.

The challenge is then to implement an execution pipeline that is compatible with the existing production infrastructure. MTIA (and all future hardware versions) are deployed in a productionized Linux container ecosystem~\citep{tupperware}. Typical production workflows require an end-to-end workload that can e.g. serve or train a model for a particular service at scale, which also requires a complicated kernel registration stack compatible with the PyTorch ecosystem. However, the Triton JIT allows us to generate, compile, and test kernels on the fly, even within these productionized containers, allowing us to run numerous experiments in parallel.

%% file: sections/03_system_design.tex
\section{System Design}
\label{sec:system_design}


Here, we describe the system architecture behind \agentname, along with the test harness used to validate the results during generation.

\subsection{In-Context Distillation of Triton MTIA Semantics}
\label{sec:philosophy}

A naive approach to generating Triton MTIA code is to add comprehensive specifications of the hardware requirements and corresponding Triton semantics differences to the first prompt of the LLM to see if the generated code passes relevant tests. In practice, however, comprehensive accelerator documentation lags behind other stack components. Early attempts at generating Triton for MTIA with a simple prompt-engineering approach resulted in significant manual labor and did not scale.\footnote{\hyperlink{https://rocm.blogs.amd.com/software-tools-optimization/triton-kernel-ai/README.html}{See zero-shot results for GEAK}.}

In contrast, \agentname{} effectively performs in-context learning iteratively, distilling hardware requirements and their corresponding Triton semantics based on the feedback obtained directly via \emph{interaction with the linter, compiler, and debugger}.

We implemented \agentname{} as a finite-state machine (FSM) with dedicated tools and routines, including linting, compiling, testing, debugging, and LLM calls.

Although recent kernel generation frameworks often rely on a dedicated reasoning agent with prescribed tool calling~\citep{sakana2025robustkbench,wang2025geak,chen2025cudallm,andrews2025gpukernel,li2025tritonbench,li2025autotriton}, we found it easier to integrate an FSM architecture within our production infrastructure \emph{at scale}. The FSM offers explicit guardrails around what is executed and performed, and allows faster debugging of the key components of the system which is an important requirement when dealing with production-ready systems. Additionally, the backend of the FSM enables compilation and testing on both the deployed MTIA machines and on the QEMU simulator of the future MTIA generations.

\subsection{\agentname~Agent}
\label{sec:agent}

\begin{figure}[ht]
    \centering
    \includegraphics[width=\linewidth]{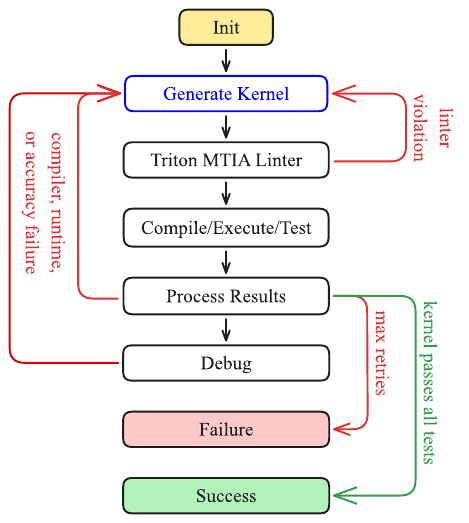}
    \caption{Finite State Machine of our kernel-generation agent \agentname. Proposal kernel-wrapper pairs are only generated during the ``Generate Kernel'' state (which dispatches to an open source, reasoning LLM). All other states process the result and update a feedback prompt, where needed.}
    \label{fig:agent_overview}
\end{figure}

 \figref{fig:agent_overview} illustrates the overall design of the \agentname{} system. Each operator is generated in a self-contained session during which all prescribed tests are performed. This session is configurable up front, allowing us to easily prototype different LLM models, disable/enable individual states (like the linter), and sweep \agentname{} hyperparameters (e.g., max number of iterations).

The routine begins with an initial prompt consisting of the task description, output requirements, the documentation (docstring) of the PyTorch operator, and three handcrafted examples~(\appenref{app:prompts}). Specifically, the prompt asks the model to generate a python wrapper matching the designated PyTorch operator signature (described in the docstring) and one or more Triton kernels that implement the functionality itself. The \texttt{exp}, \texttt{argmax}, and \texttt{diag} operators were chosen as simple examples as they span multiple kernel classes (e.g., elementwise, reduction). Often times, ATen docstrings will reference the docstrings of other operators (e.g., \texttt{argmax} references \texttt{max}). We built a directed acyclic graph of all docstrings, allowing us to include ``nested'' docstrings for completeness. Further instructions prescribing the expected output format expected by the downstream parser are also provided.

The reasoning LLM uses the prompt to generate an initial candidate wrapper/kernel pair, which is then sent to a custom Triton MTIA Linter. The linter is responsible for the following tasks: (1) ensuring the output wrapper and kernel code is compatible with the Triton JIT harness; (2) ensuring the provided implementation does not ``cheat'' by dispatching into other operators that may not yet be implemented; (3) ensuring the provided code uses valid Triton MTIA syntax and libraries, as not all of upstream Triton is available on MTIA. The linter is lightweight and configurable~(\appenref{app:linter}). If a lint violation is detected, a structured report is generated and sent back to the model as feedback for correction~(\appenref{app:prompts}). The process is repeated until no lint errors are produced or the maximum number of LLM calls is reached.

If no linter errors are detected, the wrapper and kernel code is passed to a dedicated Triton JIT compilation harness compatible with the MTIA infrastructure. Depending on the operator configuration, which prescribes the supported datatypes, a series of tests derived from  and production-data are synthesized. The test runner loops through each test, recompiling as needed (e.g. for new datatypes). If compilation is successful and the test executes without any runtime errors, the same inputs are moved to the host and executed using a reference ATen CPU implementation of the operator. The outputs for both the generated MTIA kernel and the CPU reference kernel are compared using a heuristic that depends on the underlying datatype. If the results are within the specified tolerance, the process repeats with the next test. As soon as the runner encounters a compilation failure, a runtime error, or an accuracy error, the routine breaks and proceeds to a ``feedback'' state responsible for determining what to do next.

The feedback state analyzes how successful the test runner was and determines what kind of feedback prompt is needed for another LLM iteration, or if further debugging is needed. For example, if the most recent run resulted in a runtime crash which produced a crash dump, the crash dump is loaded in an LLDB-based debugger. The debugger pulls basic information about the backtrace, decoded registers, and other frame information to provide as context for the revised prompt. Example insights include details around memory access violations.

In the case of a compiler failure, depending on the hyperparameter config, we optionally summarize the compiler log using a secondary LLM instance (also configurable up front). Triton MTIA compiler logs can easily consume thousands of tokens, so surfacing the most relevant facets of the compiler error to the main LLM session serves towards managing limited context windows.

If the feedback state detects an accuracy error, a summary of the MTIA output tensor(s) and the CPU output tensor(s) is included in the feedback prompt. Even in the case of large output tensors, an abbreviated summary of the tensor values is often enough context for the model to reason about the potential inaccuracy~(\appenref{app:reasoning}).

Once an appropriate feedback prompt is crafted, the process repeats until one of the following conditions occurs: (1) all tests pass, in which case the routine exits successfully; (2) the maximum number of prescribed LLM calls has been reached, and the routine exits; (3) the LLM context window saturates, and a new LLM dialog session starts using the most recent wrapper/kernel generation as an initial proposal; (4) an unexpected error crashes the main process. We implemented comprehensive exception handling throughout the \agentname, including launching containerized subprocesses where necessary, to avoid crashing the main process whenever possible.

In order to generate operators at scale, the above process can be executed in a parallel fashion for every operator specified and even repeated for operators that failed. These large-scale runs are configured by the operators of interest, the desired datatypes, the LLM parameters (e.g., model, context length, temperature), the run parameters (e.g., maximum number of LLM calls, maximum number of dialog sessions) and the testing complexity.

Importantly, the operators are compiled and executed on productionized MTIA machines. The LLM calls themselves are processed by a centralized inference platform service capable of handling a high volume of requests needed for large-scale runs.

\subsection{\aten{} Operators and \opinfo{} Testing}
\label{sec:opinfo}

To evaluate the viability of our approach at scale, we generate kernels for the operators defined within the PyTorch \opinfo{} testsuite. Importantly, \opinfo{} aims to rigorously test an operator's coverage by providing ``samples'' for all the supported data-types, tensor shape, and input arguments. For example, \opinfo{} contains \emph{hundreds} of tests for the \texttt{linalg.vector\_norm} operator which rigorously test different input and output tensor shapes, along with different input argument configurations.

Using \opinfo{} as our primary test harness creates an end-to-end generation pipeline with orders of magnitude more tests than the state of the art~\citep{kernelbench}. By covering more of the input space during the generation process, we expect the resulting implementation to more reliably work in arbitrary prototyping and production environments. That being said, we recognize that ensuring perfect test coverage across the entire input space is impossible. To account for this, we introduce a secondary test harness specifically for production models that consists of production input data. This additional testsuite allows us to gauge how well an operator was generalized using the \opinfo{} testsuite. If gaps are identified during the generation stage (such that production-data tests failed), then \agentname{} is able to resolve the coverage gap.

There are certain limitations with MTIA hardware such that certain operators and tests are either not compatible or not relevant for the target workloads. For example, MTIA does not support complex numbers, so we remove those corresponding operators from the generation list (e.g., FFT operators). Similarly, validating the outputs for \emph{random-number operators} between the device and host is particularly challenging due to differences in the underlying random number generation algorithm. As such, we also remove these operators from consideration. Additionally, due to a limitation of our distributed testing infrastructure, we only cover operators with under 900 total \opinfo{} tests. The resulting operator list consists of 568 unique operators (filtered down from 629).
 We also only test for \texttt{bfloat16}, \texttt{float16}, \texttt{float32}, \texttt{int32}, and \texttt{int64}. In total, this results in over \totaltests{} tests across all operators, and we only classify an operator as successful if it passes all the corresponding operator tests.

%% file: sections/04_experiments_results.tex
\section{Experiments and Results}
\label{sec:experiments_results}

We now present the experiments used to validate our approach. We first present an aggregate result consisting of kernels that span all MTIA-compatible \opinfo{} operators over multiple large-scale runs. From this set of generated operators, we ``productionize'' various first- and third-party models. We further expand our test harness for these operators by incorporating additional correctness tests that leverage \emph{production data} and identify additional gaps not originally captured by \opinfo. Finally, we ablate over various \agentname{} configurations to highlight which aspects of our pipeline matter and why.

Our baseline setup for these experiments is to run \agentname{} over all the 568 MTIA-compatible \opinfo{} operators with the following configuration:
\begin{itemize}
    \item Maximum of $3$ \agentname{} attempts (i.e., dialog sessions) per operator to generate a kernel that passes all tests and declares Success;
    \item Each attempt is allowed a maximum of $15$ LLM calls, or, in other words, $15$ full iterations through the state machine until Failure is declared for the attempt;
    \item Either Code World Model (CWM, \cite{faircodegenteam2025cwmopenweightsllmresearch}) or GPT-OSS 120B~\citep{openai2025gptoss120bgptoss20bmodel} were used as the kernel-generating LLM. Both models were configured with a context length of $131,072$ and temperature set to $1.0$. We set the top-P to $0.95$ for CWM and $1.0$ for GPT-OSS. The GPT-OSS reasoning was set to ``high.''
    \item Llama-4-Maverick is used as the feedback summarization model with the same generation parameters as CWM.
\end{itemize}

We dispatch the generation jobs across 200 production MTIA devices, which are able to finish 95\% of a run in 2 hours. The remaining tail often results from e.g., poor reasoning trajectories, and can take another 6-8 hours to complete. New runs can be dispatched concurrently. With this infrastructure in place, we executed multiple runs, with subsequent runs focusing on operators that failed previous runs.

\begin{figure}[h!]
    \centering
    \includegraphics[width=1.0\linewidth]{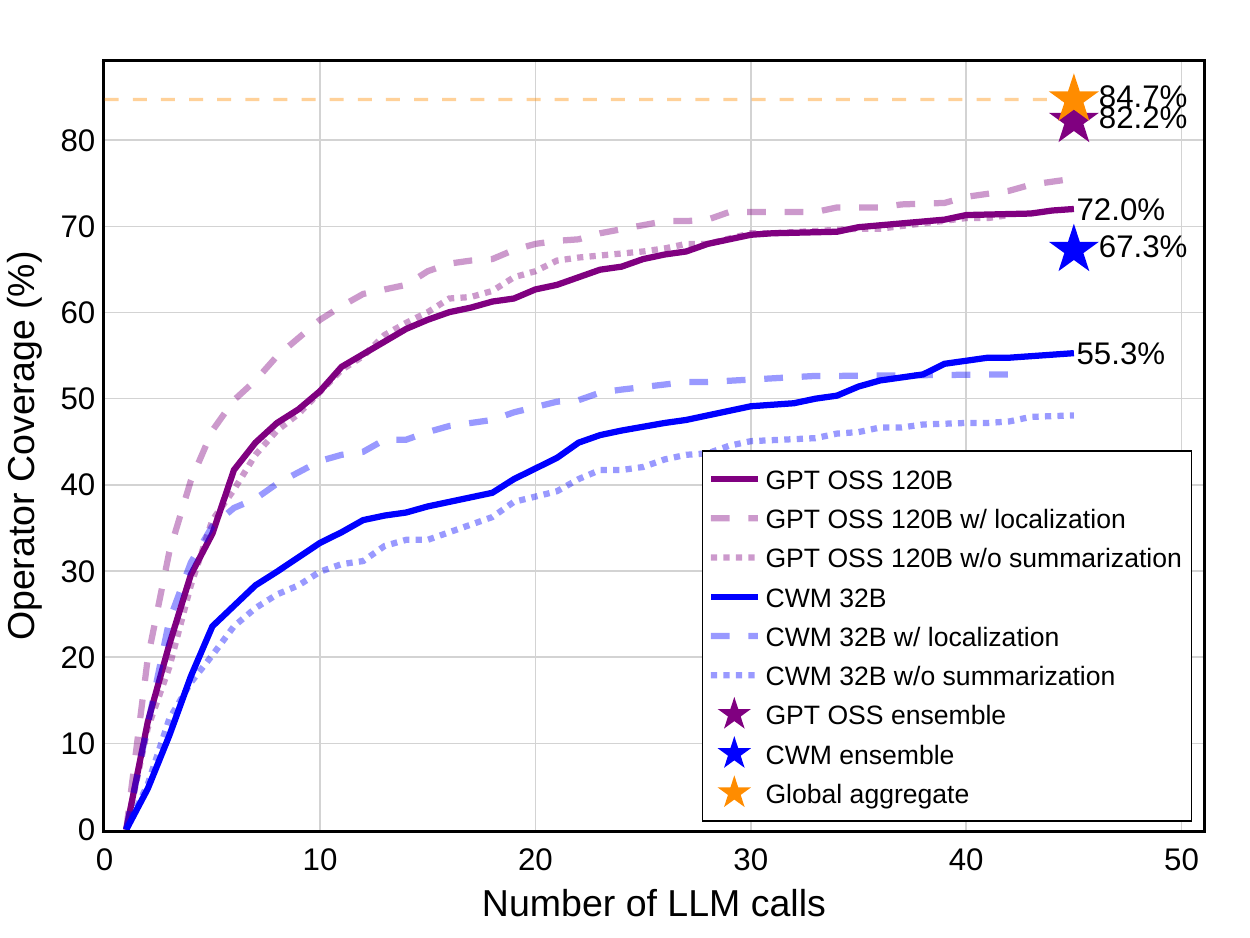}
    \caption{Number of LLM calls per operator to produce a correct kernel, cumulatively plotted for different harness configurations and models. Ensemble results display coverage achieved by combining multiple configurations. We also include experimental localization runs, where we pull relevant operators as context. The global aggregate includes all our available runs, some of which are not shown in the figure.}
    \label{fig:state_transitions}
\end{figure}

From these aggregated runs, we achieved $84.7\%$ operator coverage on all MTIA-compatible \opinfo{} operators. Here we consider an operator covered if the generated kernel-wrapper pair passes $100\%$ of the sample \opinfo{} tests. These results were aggregated across multiple runs, including ablations over models and configuration parameters. \figref{fig:state_transitions} illustrates the cumulative operator coverage as a function of LLM calls for different configurations.

 We further heuristically divide operators into $7$ categories depending on the intended functionality of each operator. Table \ref{tab:op_by_cat} shows that different categories present different difficulties to \agentname{}: while \agentname{} achieves $96.0\%$ coverage on Shape Manipulation operators, the coverage significantly drops for the operators from the Deep Learning Category.

\begin{table}[ht]
\centering
\small
\begin{tabular}{l rrr}
\toprule
 &  & \multicolumn{2}{c}{\textbf{Operator Coverage (\%)}} \\
\textbf{Op Category} & \textbf{Op Count} & \textbf{CWM} & \textbf{GPT-OSS} \\
\midrule
Elementwise       & 161 & 80.1 & 84.6 \\
Deep Learning         & 90 & 64.4 & 71.1 \\
Linear Algebra        & 78  & 71.8 & 79.5\\
Other           & 78 & 75.6 & 74.3\\
Shape Manipulation          & 75 & 96.0 & 96.0 \\
Reduction           & 63 & 69.8 & 74.6 \\
Indexing \& Selection         & 34 & 73.5 & 79.4\\
\bottomrule
\end{tabular}
\caption{\agentname{} Coverage by operator category and used LLM.}
\label{tab:op_by_cat}
\end{table}

Finally, we executed a run with GPT-OSS on a future generation using a QEMU simulator \citep{bellard2005qemu} for execution feedback. This single run yielded a coverage of $73.1\%$. We aggregated the compiler failures and feature gaps encountered during generation, and shared this data with our compiler and ASIC engineers.

\subsection{End-to-end Application}

From our baseline set of \opinfo{} operators, we used \agentname{} to enable various first- and third-party models on MTIA. While robust, we recognize that production workflows may contain operators and operator arguments (eg. shapes, scalar values, etc.) outside of the distribution represented in the \opinfo{} testsuite. To mitigate this, we decompose various target models into their individual operators, extract all operator inputs observed during training, and then use these inputs within the \agentname{} validation loop instead of those generated by \opinfo{}.

Concretely, we instrumented forward and backward passes of several representative models, NanoGPT \citep{karpathy2023nanogpt}, DLRM \citep{meta_dlrm}, and two internal recommendation models (denoted MM), using \verb|__torch_dispatch__| to intercept and record the tensor and scalar data passed to each operator. All four models were evaluated with a fixed batch size of 1024 and trained for single iteration, with the latter three (DLRM, MM1, and MM2) executed using real production data rather than randomized inputs.

Additionally, we introduce a new step to \agentname{}, first matching a given operator with a pre-generated \opinfo{} operator (should it exist), then immediately testing it with the inputs gathered from the full e2e run. Should the kernel not pass all tests out of the box, it is used as a starting point from which \agentname{}  then refines (Table~\ref{table:e2e_models}, column B: MIS).

\begin{table}[ht]
\centering
\small
\begin{tabular}{lcccc}
\toprule
 & \multicolumn{3}{c}{\textbf{Operator Coverage (\%)}} \\
 & \textbf{A. Full Model Op Set} & \multicolumn{2}{c}{\textbf{B. OpInfo Subset}} \\
\textbf{Model} &  & \textit{OpInfo} & \textit{MIS} \\
\midrule
\textbf{NGPT} & 87.2 & 80.0 & 100.0 \\
\textbf{DLRM} & 81.4 & 80.0 & 90.0 \\
\textbf{Meta M1} & 79.8 & 83.8 & 91.9 \\
\textbf{Meta M2} & 80.6 & 81.7 & 87.3 \\
\bottomrule
\end{tabular}
\caption{
Operator coverage across four model types: Nano-GPT, Deep Learning Recommendation Model, Meta Model 1, Meta Model 2. An operator is considered covered if the corresponding kernel passes all tests with model input shapes (MIS).
(A) We run \agentname{} on all model operators with the MIS feedback.  (B) We run \agentname{} on a subset of model operators that have tests available in the OpInfo suite.  \textit{(OpInfo)} We directly test kernels created with OpInfo feedback with MIS. \textit{(MIS)} We run \agentname{} to refine kernels created with OpInfo feedback with MIS feedback.
}
\label{table:e2e_models}
\end{table}

Across these experiments, \agentname{} achieves high kernel coverage, enabling nearly 80\% of all kernels required to execute a model end-to-end. Furthermore, for operators where a pre-existing \opinfo{}-validated kernel is available, over 80\% of these kernels pass all end-to-end production tests without additional prompting. After refinement, \agentname{} further improves this by an additional 6–20\% across models. This not only underscores the robustness of \agentname, but also establishes a sandbox for continuous testing and optimization of production-ready kernels.

\subsection{\agentname~Harness Ablation}
\label{sec:harness_ablations}

To  better understand which aspects of \agentname{} contribute to its success, we ablate over various configurations. Table \ref{tab:tritorx_ablations} summarizes the results of these experiments.

\begin{table}[ht]
\centering
\small
\begin{tabular}{l rr}
\toprule
\textbf{Method} & \textbf{CWM (\%)} & \textbf{GPT-OSS (\%)} \\
\midrule
Baseline (single run)       & 55.3 & 72.0\\
w/o linter                  & 48.9 & 68.7\\
w/o summarization           & 48.2 & 71.5\\
\bottomrule
\end{tabular}
\caption{Ablations over \agentname{} harness features.}
\label{tab:tritorx_ablations}
\end{table}

We examine the importance of the custom Triton MTIA linter and the optional summarization model. Removing the linter resulted in a significant drop in performance ($55.3\%\to48.9\%$ for CWM). As mentioned previously, the linter not only helps the agent identify intrinsics unique to the Triton MTIA dialect, but also helps prevent ``cheating'' by flagging the unauthorized use of other torch operators (\appenref{app:linter}).

Removing the summarization agent also resulted in a decrease in performance ($55.3\%\to48.2\%$ for CWM). Without a separate summarization agent, the entire compilation log, which can consist of thousands of tokens, is fed directly into the LLM dialog session, and the model performance can degrade as we approach the context limit \citep{hsieh2024rulerwhatsrealcontext}.

%% file: sections/06_related_work.tex
\section{Related Work}
\label{sec:related_work}

\textbf{Custom ASICs, MTIA \& Deep Learning Compilers.}
Since Triton introduced a Python-first DSL for high-performance GPU kernels~\citep{tillet2019triton} and became widely used by PyTorch Inductor~\citep{ansel2024pytorch} to generate fused operators, several ecosystems now offer Python-level kernel DSLs. NVIDIA's Warp~\citep{warp2022} is a Python DSL for authoring CUDA kernels, with an optional tile-based programming model for tensor-core GEMMs. In JAX, Google's Pallas~\citep{jax2018github} allows users to write custom kernels in Python; it targets GPUs via Triton and TPUs~\citep{jouppi2017datacenter} via Mosaic~\citep{bansal2023mosaic}, enabling fine-grained fused operations within the JAX ecosystem. Meta has also adopted Triton for MTIA to improve developer efficiency with PyTorch.

\textbf{Kernel Research.}
Optimization of custom kernels for specific devices has received sustained attention in recent years: ~\cite{lavin2016fast,dao2022flashattention,dao2023flashattention2,shah2024flashattention3}.
This interest is both driven by and enabling the rapid growth of training and inference compute of frontier models:~\cite{epoch2024computetrend, hooker2021hardware}.

\textbf{Benchmarks.}
The community has responded by releasing benchmarks to measure the functional correctness and performance of LLM-generated kernels that primarily target GPUs through CUDA and Triton.
\textsc{KernelBench}~\citep{kernelbench} and \textsc{TritonBench}~\citep{li2025tritonbench} evaluate whether models can generate correct and performant GPU kernels across representative operator suites, while \textsc{NPU-Eval}~\citep{kalade2025npueval} targets AMD NPUs. \textsc{AlgoTune}~\citep{press2025algotune} in particular targets LLM's ability to speed up scientific computing problems from natural language descriptions.   
Automatically generated kernels are prone to exploiting holes in the test suite. Addressing this has seen significant effort:~\citep{sakana2025robustkbench, metr2025measuring}. Notably, similar to our work \textsc{BackendBench}~\citep{saroufim2025backendbench} also uses PyTorch's OpInfo as a comprehensive test suite to ensure correctness and comprehensiveness.

\textbf{LLM-based Kernel Generation.}
Most LLM-based kernel generation relies on prompting techniques combined with test-time compute use as a form of search.
Recent systems train or prompt LLMs to synthesize kernels or improve them with search.
\textsc{KernelLLM}~\citep{kernelllm2025} provides a supervised baseline for the generation of Triton kernels from PyTorch modules. \textsc{AutoTriton}~\citep{li2025autotriton} adds reinforcement learning from verifiable rewards. Multi-turn reinforcement learning for CUDA kernel generation has also been explored in~\cite{baronio2025kevin}.

Orthogonally, several approaches scale inference compute with existing LLMs, covering a spectrum of prompting, agentic, and evolutionary techniques combined with verification through unit tests:~\citep{sakana2025robustkbench,metr2025measuring,nvidia2025deepseekr1,wei2025astra,wang2025geak}.
While several approaches exist that target CUDA and GPUs, we believe only NPU-Eval~\citep{kalade2025npueval} is targeting non-GPU devices with C++.

%% file: sections/07_discussion.tex
\section{Discussion and Future Work}
\label{sec:discussion}
\agentname's FSM framework provides a robust harness for generating functionally correct Triton MTIA kernels. Thanks to its flexibility, we identify several additional directions to further improve coverage and performance and reflect on what aspects are most important for success.

\textbf{Self-consistent operator generation.} To prevent ``cheating,'' the linter restricts the agent from utilizing other \aten{} operators within the wrapper (beyond tensor allocation). A more efficient and possibly more performant approach is to allow the wrapper to dispatch other operators, provided they are also implemented in the new backend and the operators do not result in cyclic dependencies. This requires a self-consistent generation scheme where the agent is aware of the entire backend state (and is thus no longer embarrassingly parallel).

\textbf{Optimized prompting.} As mentioned earlier, we found that simpler prompts without dedicated MTIA documentation worked best, but there remains room for prompt tuning. Furthermore, we can improve the quality of the example kernel-wrapper pairs themselves using strategies like localization, perhaps even bootstrapping the process over sequential runs.

\textbf{Fully agentic pipeline.} At a higher level than these improvements, we have discussed making the entire FSM agentic -- converting most current states into tools for the LLM to call upon, as well as adding new tools related to debugging, such as giving the LLM a sandbox to execute code in.

\textbf{Dedicated model post-training.} Both models used throughout this work were open source and off-the-shelf. All of the MTIA-specific context was gained via interactions with the linter, compiler, and debugger. An orthogonal research direction would be to further post-train the LLM for Triton MTIA kernel generation and contrast that with the existing results for vanilla Triton in the literature.

\textbf{The importance of scale.} Due to the stochastic nature of the underlying language model, running the benchmark repeatedly (with a nonzero temperature) will produce results with nonidentical pass-rates. Thus, simply aggregating the passing operators across runs, a technique known as test-time scaling, can yield significant increases in operator coverage. For example, just aggregating between two benchmark runs using CWM increased the coverage from $55\%\to64\%$. We suspect that further scaling this and exploring more complex strategies such as evolutionary scaling ~\citep{sakana2025robustkbench} will bring significant coverage improvements. Furthermore, supporting additional hardware generations, DSLs, and operator definitions will require a flexible and scalable pipeline for generating operator sets.

\textbf{Establishing a backend-maintenance environment.} As AI-driven kernel generation evolves, AI will eventually produce more kernels than humans can feasibly review—especially as hardware diversity increases and multiple generations of hardware coexist within production fleets. Establishing a robust, fully automated framework that does not require human review will be essential for the success of AI-generated kernels. This need will become even more critical as each model update triggers a comprehensive refresh of the kernel library to optimize performance.

%% file: sections/08_conclusion.tex
\section{Conclusion}
\label{sec:conclusion}

We introduced \agentname, a scalable, coverage-first system that automates bring-up of the PyTorch backend for custom ML ASICs, demonstrated on Meta’s MTIA. \agentname{} orchestrates an existing LLM with a finite-state workflow and production-compatible tooling, executed on both deployed silicon and a QEMU-based simulator for future devices. It generated over \textbf{481} \aten{} operators that pass all of their corresponding \opinfo{} tests (\textbf{\totaltests{}+} total tests), achieving an overall pass rate of \textbf{84.7\%}. From these operators, we were able to onboard \textbf{80+\%} of the multiple first- and second-party large-scale models on the device. Ablations over model and system factors isolate which components most influence coverage, turning backend enablement into a measured engineering process rather than an artisanal effort.

Beyond MTIA, the design is applicable to other accelerators and naturally extends to adjacent tasks such as cross-generation kernel migration and early hardware/compiler feedback via simulation. Although performance tuning is out of the scope of this work, \agentname{} provides the substrate on which autotuning, schedule search, and learned code optimization can be layered. Future directions include expanding operator families (reductions, sparse/quantized ops), stronger safety/containment, and formal checks for code generation, tighter integration with compiler IRs, and automated pathways for performance refinement.

Taken together, these results offer a practical blueprint for democratizing toolchain creation and a force multiplier for kernel engineers—freeing expert effort to focus on truly performance-critical paths while bringing entire backends online in hours rather than months.

%% file: sections/99_appendix.tex
\renewcommand\thefigure{\thesection.\arabic{figure}}

\section{Acknowledgments}
We thank the PyTorch Team, in particular the authors of BackendBench for their support. We acknowledge that our work stands on the shoulders of the entire MTIA team and are grateful for the enablement of our research. We also thank Dipal Saluja for his early feedback.

\section{Generated Wrapper and Kernel Pairs}\label{app:example_kernels}
Here we present a few interesting kernel-wrapper pairs for different operators.

\subsection{nn.functional.binary\_cross\_entropy}

\begin{lstlisting}[
style=pythonstyle,    basicstyle=\scriptsize\ttfamily,
numbers=none
]
@triton.jit
def kernel(
    input_ptr,
    target_ptr,
    weight_ptr,
    output_ptr,
    n_elements,
    BLOCK_SIZE: tl.constexpr,
):
    pid = tl.program_id(axis=0)
    block_start = pid * BLOCK_SIZE
    offsets = block_start + tl.arange(0, BLOCK_SIZE)
    mask = offsets < n_elements
    input = tl.load(input_ptr + offsets, mask=mask, other=0.0)
    target = tl.load(target_ptr + offsets, mask=mask, other=0.0)
    weight = tl.load(weight_ptr + offsets, mask=mask, other=1.0)
    eps = 1e-8
    loss = -target * tl.log(input + eps) - (1 - target) * tl.log(1 - input + eps)
    loss = loss * weight
    tl.store(output_ptr + offsets, loss, mask=mask)

def wrapper(input, target, weight=None, reduction='mean'):
    if input.shape != target.shape:
        raise RuntimeError("input and target must have the same shape")
    if weight is not None:
        if weight.shape != input.shape:
            weight = weight.broadcast_to(input.shape)
        weight = weight.contiguous()
    else:
        weight = torch.ones_like(input, dtype=input.dtype, device=input.device)
    if reduction not in ['none', 'mean', 'sum']:
        raise ValueError("reduction must be 'none', 'mean', or 'sum'")
    output = torch.empty_like(input)
    input = input.contiguous()
    target = target.contiguous()
    n_elements = input.numel()
    if n_elements == 0:
        if reduction == 'none':
            return output
        else:
            return torch.tensor(0.0, dtype=input.dtype, device=input.device)
    BLOCK_SIZE = 1024
    grid = (triton.cdiv(n_elements, BLOCK_SIZE),)
    kernel[grid](
        input,
        target,
        weight,
        output,
        n_elements,
        BLOCK_SIZE=BLOCK_SIZE,
    )
    if reduction == 'none':
        return output
    elif reduction == 'sum':
        return output.sum()
    elif reduction == 'mean':
        return output.mean()
\end{lstlisting}

\newpage
\subsection{outer}

\begin{lstlisting}[
style=pythonstyle,    basicstyle=\scriptsize\ttfamily,
numbers=none
]
@triton.jit
def kernel(
    input_ptr,
    vec2_ptr,
    output_ptr,
    n,
    m,
    input_stride,
    vec2_stride,
    output_stride0,
    output_stride1,
):
    pid = tl.program_id(0)
    if pid >= n:
        return
    input_val = tl.load(input_ptr + pid * input_stride)
    for j in range(m):
        vec2_val = tl.load(vec2_ptr + j * vec2_stride)
        product = input_val * vec2_val
        output_offset = pid * output_stride0 + j * output_stride1
        tl.store(output_ptr + output_offset, product)

def wrapper(input, vec2, *, out=None):
    if input.dim() != 1 or vec2.dim() != 1:
        raise ValueError("Both input and vec2 must be 1D tensors")
    n = input.size(0)
    m = vec2.size(0)
    output_shape = (n, m)
    dtype = input.dtype
    if input.dtype != vec2.dtype:
        pass
    device = input.device
    if out is None:
        output = torch.empty(output_shape, dtype=dtype, device=device)
    else:
        if out.shape != output_shape:
            raise RuntimeError(f"Expected out tensor to have shape {output_shape}, but got {out.shape}")
        if out.dtype != dtype:
            raise RuntimeError(f"Expected out tensor to have dtype {dtype}, but got {out.dtype}")
        if out.device != device:
            raise RuntimeError(f"Expected out tensor to be on device {device}, but got {out.device}")
        output = out
    input_contig = input.contiguous()
    vec2_contig = vec2.contiguous()
    output_contig = output.contiguous()
    input_stride = input_contig.stride(0)
    vec2_stride = vec2_contig.stride(0)
    output_stride0 = output_contig.stride(0)
    output_stride1 = output_contig.stride(1)
    grid = (n,)
    kernel[grid](
        input_contig,
        vec2_contig,
        output_contig,
        n,
        m,
        input_stride,
        vec2_stride,
        output_stride0,
        output_stride1,
    )
    return output
\end{lstlisting}

\newpage
\subsection{nn.functional.layer\_norm}
\begin{lstlisting}[
style=pythonstyle,    basicstyle=\scriptsize\ttfamily,
numbers=none
]
@triton.jit
def kernel_mean_var(
    input_ptr,
    mean_ptr,
    var_ptr,
    M,
    N,
    epsilon,
    BLOCK_SIZE: tl.constexpr,
):
    pid = tl.program_id(0)
    if pid >= N:
        return

    sum = 0.0
    sum_sq = 0.0
    for i in range(M):
        x = tl.load(input_ptr + pid * M + i)
        sum += x
        sum_sq += x * x

    mean = sum / M
    var = sum_sq / M - mean * mean + epsilon

    tl.store(mean_ptr + pid, mean)
    tl.store(var_ptr + pid, var)
@triton.jit
def kernel_normalize(
    input_ptr,
    output_ptr,
    mean_ptr,
    var_ptr,
    weight_ptr,
    bias_ptr,
    M,
    N,
    epsilon,
    elementwise_affine,
    BLOCK_SIZE: tl.constexpr,
):
    pid = tl.program_id(0)
    if pid >= N:
        return

    mean = tl.load(mean_ptr + pid)
    var = tl.load(var_ptr + pid)

    for i in range(M):
        x = tl.load(input_ptr + pid * M + i)
        x_float = x.to(tl.float32)
        x_float = x_float - mean
        x_float = x_float / tl.sqrt(var)  # Corrected line: removed epsilon here

        if elementwise_affine != 0:
            weight = tl.load(weight_ptr + i).to(tl.float32)
            bias = tl.load(bias_ptr + i).to(tl.float32)
            x_float = x_float * weight + bias

        x = x_float.to(x.dtype)
        tl.store(output_ptr + pid * M + i, x)

def wrapper(input, normalized_shape, weight=None, bias=None, eps=1e-5, elementwise_affine=True):
    if not isinstance(input, torch.Tensor):
        raise TypeError("Input must be a torch.Tensor")
    if not input.is_contiguous():
        input = input.contiguous()

    if isinstance(normalized_shape, int):
        normalized_shape = (normalized_shape,)
    else:
        normalized_shape = tuple(normalized_shape)
    D = len(normalized_shape)
    input_shape = input.shape
    if D > len(input_shape):
        raise ValueError("normalized_shape cannot be larger than input shape")

    M = 1
    for dim in normalized_shape:
        M *= dim
    N = 1
    for dim in input_shape[:-D]:
        N *= dim

    if N == 0 or M == 0:
        return torch.empty_like(input)

    input_float = input.to(torch.float32)
    input_float = input_float.view(N, M)
    mean = torch.empty(N, dtype=torch.float32, device=input.device)
    var = torch.empty(N, dtype=torch.float32, device=input.device)

    BLOCK_SIZE = 1
    grid = (N,)
    kernel_mean_var[grid](
        input_float,
        mean,
        var,
        M,
        N,
        eps,
        BLOCK_SIZE=BLOCK_SIZE,
    )

    output = torch.empty_like(input)
    output = output.view(N, M)

    if elementwise_affine:
        if weight is None:
            weight = torch.ones(normalized_shape, dtype=input.dtype, device=input.device)
        else:
            weight = weight.contiguous().view(-1)
        if bias is None:
            bias = torch.zeros(normalized_shape, dtype=input.dtype, device=input.device)
        else:
            bias = bias.contiguous().view(-1)
    else:
        weight = torch.ones(M, dtype=input.dtype, device=input.device)
        bias = torch.zeros(M, dtype=input.dtype, device=input.device)

    input_flattened = input.view(N, M)
    kernel_normalize[grid](
        input_flattened,
        output,
        mean,
        var,
        weight,
        bias,
        M,
        N,
        eps,
        1 if elementwise_affine else 0,
        BLOCK_SIZE=BLOCK_SIZE,
    )

    output = output.view(input_shape)
    return output
\end{lstlisting}

\newpage
\section{State Transition Prompts}\label{app:prompts}








As \agentname{} executes its feedback loop, we have different prompts for different scenarios. The initial prompt provides rules and background information, most critically PyTorch docstring(s) explaining both the operator in question and other referenced PyTorch docs or operators, alongside three fully working MTIA Triton reference kernels for \texttt{exp}, \texttt{argmax}, and \texttt{diag}:

\begin{lstlisting}[
style=pythonstyle,    basicstyle=\scriptsize\ttfamily,
numbers=none
]
init_prompt = f"""
Write me a Triton implementation of the {op_name} ATen operator from PyTorch using MTIA's version of Triton. This implementation should support the following input dtypes: {str(dtypes)}.

To make it easy to use for me please make sure to call your jitted kernel "kernel" and the wrapper code that translates the original signature of {op_name} to the kernel's signature "wrapper".

Multiple kernels are fine if needed, but try to generalize as much as possible and limit the number of kernels. Thoroughly analyze the problem and decide if one kernel is enough or if multiple are needed. If you get stuck trying to put all logic in one kernel, feel free to create multiple. Note that if you have multiple kernels, their names MUST all start with "kernel".

Do NOT use or fallback to the base PyTorch or Triton implementations of these kernels - this is considered cheating and is NOT allowed.

Do NOT include any comments or import statements (the required ones will be added while running the kernel - do not worry about import errors).

Additionally, for MTIA, we add 32-bit padding to input tensors. Take this into account while you are iterating and debugging your kernel.

Think carefully about a correct and efficient implementation and finally output your code in a Python codeblock. Please only output a single module's code containing the kernel and wrapper functions.

I'll paste the docstr of ATen's {op_name} for reference, which defines the spec:
{docstring}
{supplemental_docstrings}

\n\n\nFor your reference, I am including a few different types of fully working MTIA Triton implementations of ATen operators.

These will be listed in triplets, where each example will have "Operator", "Kernel(s)", and "Wrapper". Analyze these to better understand MTIA Triton:\n

<reference kernels>

Please think carefully and output a full implementation of {op_name} in MTIA Triton now!"""
\end{lstlisting}

If we are starting a new LLM session but have a partially working kernel already, we instead prompt it to \textit{debug} an existing kernel, and provide the current implementation:

\begin{lstlisting}[
style=pythonstyle,    basicstyle=\scriptsize\ttfamily,
numbers=none
]
init_prompt_with_existing_kernel = f"""
Debug a Triton implementation of the {op_name} ATen operator from PyTorch, written in MTIA's version of Triton. This implementation should support the following input dtypes: {str(dtypes)}.
...
...
Lastly, here is the work-in-progress implementation of this operator. It has a few issues that I need your help debugging, so be sure to analyze it thoroughly:
<current partial implementation>
"""
\end{lstlisting}

We describe the details of linting more in Appendix \ref{app:linter}, but the prompt itself is fairly straightforward, simply showing the LLM the lint errors.

Our feedback prompts for compilation errors and correctness errors are more nuanced, and consist of three separate prompts. To begin with, part of our approach to optimize context length includes using an LLM-based summarizer for long compilation logs that often contain the same error message multiple times:

\begin{lstlisting}[
style=pythonstyle,    basicstyle=\scriptsize\ttfamily,
numbers=none
]
summarization_prompt = """
...
...
To recap, return the following information ONLY:
1. The EXACT error message
2. The EXACT code snippet that caused the error, both the exact line and the lines before it
3. The EXACT traceback of the error, if present - do NOT include duplicates
"""
\end{lstlisting}

This summary is then passed into a simple prompt telling the LLM to evaluate the compilation error and fix the kernel.

When the generated kernel fails a test case due to a correctness error, we provide quite a bit of feedback to the LLM based on the test case to use as context while debugging. We've found that this does not increase the risk of overfitting to a specific test case whilst increasing the success rate:

\lstset{aboveskip=5pt, belowskip=5pt}
\begin{lstlisting}[
style=pythonstyle,    basicstyle=\scriptsize\ttfamily,
numbers=none
]
correctness_feedback_prompt = """
...
...
**Summary of the CPU output tensor for the input**:
{str(latest_result.cpu_tensor_results)}

**Summary of the MTIA output tensor for the input**:
{str(latest_result.mtia_tensor_results)}

Additionally, here is the input and output data for the first failing test case. Do NOT overfit your kernel to this - remember you need to build a generalized kernel implementation. This is here solely to help you debug.

**INPUT SIGNATURE**:
{str(input_sig)}

**OUTPUT SIGNATURE**:
{str(output_sig)}

**INPUT SHAPE**:
{str(latest_result.shape)}

**INPUT TENSOR**:
{str(latest_result.input_tensor)}

**INPUT ARGS**:
{str(latest_result.input_args)}

**INPUT KWARGS**:
{str(latest_result.input_kwargs)}

Generate the corrected MTIA kernel now:
"""
\end{lstlisting}

\section{Reasoning Trajectories}
\label{app:reasoning}

\vspace{-0.3cm}
\begin{figure}[H]
{
\lstset{aboveskip=0pt, belowskip=5pt}
\begin{lstlisting}[style=cwm_prompt]
You are an expert in generating Triton MTIA (Meta Training and Inference Accelerator) kernels. Write me a Triton implementation of the nn.functional.logsigmoid ATen operator from PyTorch using MTIA's version of Triton. This implementation should support the following input dtypes: ['bfloat16', 'float16', 'float32'].
...
I'll paste the docstr of ATen's nn.functional.logsigmoid for reference, which defines the spec:
logsigmoid(input) -> Tensor
Applies element-wise :math:`\text{LogSigmoid}(x_i) = \log \left(\frac{1}{1 + \exp(-x_i)}\right)`
See :class:`~torch.nn.LogSigmoid` for more details.
For your reference, I am including a few different types of fully working MTIA Triton implementations of ATen operators.
...
\end{lstlisting}
\begin{lstlisting}[style=cwm_think]
Okay, I need to implement the logsigmoid function using MTIA Triton. Let's start by understanding the problem. The logsigmoid function applies the element-wise operation LogSigmoid(x_i) = log(1 / (1 + exp(-x_i))). This is equivalent to -softplus(-x_i), where softplus is the softplus activation function.

First, I should consider the input dtypes: bfloat16, float16, and float32. MTIA Triton might handle these differently, but I'll assume that the kernel can be written in a way that works for all three types, possibly using type specialization or generic operations.
...
\end{lstlisting}
\begin{lstlisting}[style=cwm_act]
@triton.jit
def kernel(
    input_ptr,
    output_ptr,
    n_elements,
    BLOCK_SIZE: tl.constexpr,
):
    pid = tl.program_id(0)
    block_start = pid * BLOCK_SIZE
    offsets = block_start + tl.arange(0, BLOCK_SIZE)
    mask = offsets < n_elements

    x = tl.load(input_ptr + offsets, mask=mask)
    exp_neg_x = tl.exp(-x)
    logsigmoid = -tl.log1p(exp_neg_x)
    tl.store(output_ptr + offsets, logsigmoid, mask=mask)

def wrapper(input):
    output = torch.empty_like(input, device=input.device)
    if input.numel() == 0:
        return output
    input_contig = input.contiguous().view(-1)
    output_contig = torch.empty_like(input_contig, device=input.device)
    n_elements = input.numel()
    BLOCK_SIZE = 128
    grid = (triton.cdiv(n_elements, BLOCK_SIZE),)
    kernel[grid](
        input_contig,
        output_contig,
        n_elements,
        BLOCK_SIZE=BLOCK_SIZE,
    )
    return output_contig.view(input.shape)
\end{lstlisting}
\begin{lstlisting}[style=cwm_obs]
Your previous MTIA kernel implementation failed to pass the linter. Please analyze the lint error and provide a corrected version. Remember to adhere to the original output guidelines.
**Lint Error**:
Found 1 linting violation(s):
[module_restrictions] Forbidden tl module usage: tl.log1p (line 15)
Details: Allowed tl functions: tl.abs, tl.add, tl.advance, ...
...
Generate the corrected MTIA kernel now
\end{lstlisting}
\begin{lstlisting}[style=cwm_think]
Okay, the linter is complaining about the use of tl.log1p in the kernel. Let me check the allowed tl functions. The error message lists the allowed functions, and tl.log1p is not among them.
...
\end{lstlisting}
\begin{lstlisting}[style=cwm_act]
...
\end{lstlisting}
\begin{lstlisting}[style=cwm_obs]
Your MTIA kernel implementation failed to compile. Please analyze the compilation error and provide a corrected version. Remember to adhere to the original output guidlines.
**Compilation Error**:
...
ValueError: Expected dtype ['fp32', 'fp64'] but got fp16
...
Generate the corrected MTIA kernel now:
\end{lstlisting}
\begin{lstlisting}[style=cwm_think]
...
\end{lstlisting}
\begin{lstlisting}[style=cwm_act]
def kernel(
    input_ptr,
    output_ptr,
    n_elements,
    BLOCK_SIZE: tl.constexpr,
):
    pid = tl.program_id(0)
    block_start = pid * BLOCK_SIZE
    offsets = block_start + tl.arange(0, BLOCK_SIZE)
    mask = offsets < n_elements

    x = tl.load(input_ptr + offsets, mask=mask)
    x_float32 = tl.cast(x, tl.float32)
    exp_neg_x = tl.exp(-x_float32)
    logsigmoid_float32 = -tl.log(1 + exp_neg_x)
    logsigmoid = tl.cast(logsigmoid_float32, x.dtype)
    tl.store(output_ptr + offsets, logsigmoid, mask=mask)

def wrapper(input):
    ...
\end{lstlisting}
}
\caption{Example of \agentname{} writing a kernel-wrapper pair for the \texttt{nn.functional.logsigmoid} operator in $3$ LLM calls after receiving the linter and compiler feedback. The reasoning steps are shortened and some details are omitted due to space constraints. Initial prompt in green, LLM reasoning in blue, LLM kernel-wrapper generation in purple, and \agentname{} feedback in orange.}
\label{fig:reas_traj_1}
\end{figure}

\begin{figure}[H]
{
\lstset{aboveskip=0pt, belowskip=5pt}
\begin{lstlisting}[style=cwm_prompt]
You are an expert in generating Triton MTIA (Meta Training and Inference Accelerator) kernels. Write me a Triton implementation of the nn.functional.channel_shuffle ATen operator from PyTorch using MTIA's version of Triton. This implementation should support the following input dtypes: ['bfloat16', 'float16', 'float32', 'int32', 'int64'].
...
I'll paste the docstr of ATen's nn.functional.channel_shuffle for reference, which defines the spec:
channel_shuffle(input, groups) -> Tensor
Divide the channels in a tensor of shape :math:`(*, C , H, W)`
into g groups and rearrange them as :math:`(*, C \frac g, g, H, W)`,
while keeping the original tensor shape.
...
\end{lstlisting}
\begin{lstlisting}[style=cwm_think]
Okay, so I need to implement the nn.functional.channel_shuffle operator using MTIA Triton. Let me start by understanding what this operator does.
...
\end{lstlisting}
\begin{lstlisting}[style=cwm_act]
@triton.jit
def kernel(
    input_ptr,
    output_ptr,
    ...
):
    ...
        values = tl.load(input_ptr + input_offsets, mask=mask)
        tl.store(output_ptr + output_offsets, values, mask=mask)

def wrapper(input, groups):
    ...
\end{lstlisting}
\begin{lstlisting}[style=cwm_obs]
Your MTIA kernel implementation failed to compile. Please analyze the compilation error and provide a corrected version. Remember to adhere to the original output guidlines.
**Compilation Error**:
Compilation log with error:
error: Scatter stores are disabled by default. Please set the "enable_scatter_stores" flag or revisit the algorithm to avoid this unsafe pattern.
error: failed to legalize operation 'tts.scatter' that was explicitly marked illegal
...
Generate the corrected MTIA kernel now
\end{lstlisting}
\begin{lstlisting}[style=cwm_think]
Okay, the compilation error is due to scatter stores being disabled by default in Triton. The error message suggests that the kernel is using a scatter store operation, which is not allowed unless explicitly enabled.
Looking at the kernel code, the issue arises from the line ... This is a scatter operation because each thread is writing to potentially different, non-contiguous memory locations based on ...
To fix this, we need to avoid scatter stores. One way to do this is to change the approach so that each thread writes to a contiguous block of memory.
...
\end{lstlisting}
\begin{lstlisting}[style=cwm_act]
...
\end{lstlisting}
\begin{lstlisting}[style=cwm_obs]
While running tests, your MTIA kernel implementation failed to produce the correct output (wrt a CPU implementation, which is our source of truth).
Note that it compiled successfully. It just failed to produce the correct output for a test case.
** Please analyze the differences between the CPU output and your MTIA kernel's output and provide a corrected version. **
**Summary of the CPU output tensor for the input**:
...
**Summary of the MTIA output tensor for the input**:
...
Generate the corrected MTIA kernel now
\end{lstlisting}
\begin{lstlisting}[style=cwm_think]
Let me analyze the differences between the expected CPU output and the MTIA kernel's output to find the issue.
First, I'll compare the input and output tensors. Looking at the MTIA output, there are zero values in some positions where the CPU output has non-zero values.
...
\end{lstlisting}
\begin{lstlisting}[style=cwm_act]
...
\end{lstlisting}
\begin{lstlisting}[style=cwm_obs]
The provided MTIA kernel implementation compiled but had a PE crash on MTIA hardware.
This is often caused by memory access errors. Please analyze the coredump and provide a corrected version.
...
\end{lstlisting}
\begin{lstlisting}[style=cwm_think]
Okay, so the MTIA kernel implementation is crashing due to a machine external interrupt, likely caused by a memory access error. The root cause is probably an out-of-bounds access or incorrect memory alignment. Let's think about possible causes.
...
\end{lstlisting}
\begin{lstlisting}[style=cwm_act]
...
\end{lstlisting}
\begin{lstlisting}[style=cwm_obs]
Your MTIA kernel implementation failed to compile. Please analyze the compilation error and provide a corrected version. Remember to adhere to the original output guidlines.
**Compilation Error**:
...
ValueError: arange's arguments must be of type tl.constexpr
...
\end{lstlisting}
\begin{lstlisting}[style=simple_small_small]
...
\end{lstlisting}
\begin{lstlisting}[style=cwm_act]
@triton.jit
def kernel(
    input_ptr,
    output_ptr,
    ...
    BLOCK_SIZE: tl.constexpr,
):
    ...

    # Process each position and group to avoid scatter stores
    for pos in range(0, k):
        for group in range(0, g):
            c_orig = group * k + pos
            c_new = pos * g + group

            input_offset = input_base + c_orig * stride_c
            output_offset = output_base + c_new * stride_c

            value = tl.load(input_ptr + input_offset)
            tl.store(output_ptr + output_offset, value)

def wrapper(input, groups):
...
\end{lstlisting}
}
\caption{Example of \agentname{} writing a kernel-wrapper pair for the \texttt{nn.functional.channel-shuffle} operator in $14$ LLM calls. The LLM has to incorporate the feedback from the \texttt{scatter-store} compilation error (hardware restriction) and the discrepancy between the MTIA test output compared to the CPU implementation, as well as address other compilation errors and hardware crashes.}
\label{fig:reas_traj_2}
\end{figure}

\newpage
\section{Triton MTIA Linter}\label{app:linter}

The linter is a rule-based, static analysis tool that combines Python's AST parsing with regex matching per rule. Several of these rules are formatting restrictions to make the output compatible with our execution system, such as disallowing import statements and enforcing a specific output format.

Other rules are more concerned with what modules are being used and where. We have allowlists on several key modules, such as \texttt{tl} (\texttt{triton.language}), and \texttt{torch}. Anything from one of these modules not found in such an allowlist is considered a linter violation:
\newline
\begin{lstlisting}[style=pythonstyle]
module_restrictions:
  modules:
    - module_name: "tl"
      allowed_functions:
        - "tl.load"
        - "tl.store"
        - "tl.arange"
        # ... 200+ allowed Triton MTIA operations
    - module_name: "torch"
      allowed_functions:
        - "torch.empty"
        - "torch.zeros"
        # ... tensor allocation/reshaping only

\end{lstlisting}

This highlights exactly which aspects of upstream Triton are compatible with MTIA.

Additionally, we have \emph{scope restrictions} on modules, such as allowing \texttt{tl} only in the kernel, not in the wrapper:
\newline
\begin{lstlisting}[style=pythonstyle]
module_scope_restrictions:
  restrictions:
    - module: "tl"
      allowed_scope_patterns: ["^kernel.*"]  # tl.* only inside kernel functions
\end{lstlisting}

As part of our anti-cheating efforts, we enforce several rules. To prevent moving tensors back to CPU or attempting to move tensors to CUDA:
\newline
\begin{lstlisting}[style=pythonstyle]
forbidden_tensor_methods:
  enabled: true
  description: "Prohibit tensor methods that move data between devices (CPU/CUDA transfers)"
  forbidden_methods:
    - "cpu"      # tensor.cpu() - moves tensor to CPU
    - "cuda"     # tensor.cuda() - moves tensor to CUDA

forbidden_function_arguments:
  enabled: true
  description: "Prohibit specific argument values in function calls"
  restrictions:
    # Forbid explicit CPU/CUDA device specifications in torch.device()
    - function: "torch.device"
      forbidden_string_args:
        - "cpu"
        - "cuda"
\end{lstlisting}

And to prevent workarounds such as \texttt{eval} or \texttt{exec}:
\newline
\begin{lstlisting}[style=pythonstyle]
# Rule to ban dangerous built-in functions that enable dynamic code execution
forbidden_functions:
  enabled: true
  description: "Prohibit built-in functions that enable dynamic code execution"
  forbidden_functions:
    - "eval"    # Evaluates Python expressions from strings
    - "exec"    # Executes Python code from strings
    - "compile" # Compiles Python code from strings (used with exec)
\end{lstlisting}